%% file: Abstract.tex
\documentclass[10pt,twoside,russian]{winter_1}
\RequirePackage[cp1251]{inputenc}
\RequirePackage[T2A]{fontenc}
\RequirePackage[russian,english]{babel} \RequirePackage{epsfig}
\RequirePackage{amsmath} \RequirePackage{amssymb}
\RequirePackage{amsfonts} \RequirePackage{longtable}
\RequirePackage{verbatim}
\usepackage{chapterbib}
\usepackage{indentfirst}
\usepackage{cite}
\usepackage{url}
\usepackage{russmath}

\usepackage{lscape}
\usepackage[small,labelsep=period]{caption}
\DeclareCaptionLabelFormat{numlesscap}{#1~}

\renewcommand\ne{\mathchar"3236\mathchar"303D\nobreak
      \discretionary{}{\usefont
      {OMS}{cmsy}{m}{n}\char"36\usefont
      {OT1}{cmr}{m}{n}\char"3D}{}}
\begingroup
\catcode`\+\active\gdef+{\mathchar8235\nobreak\discretionary{}%
 {\usefont{OT1}{cmr}{m}{n}\char43}{}}
\catcode`\-\active\gdef-{\mathchar8704\nobreak\discretionary{}%
 {\usefont{OMS}{cmsy}{m}{n}\char0}{}}
\catcode`\=\active\gdef={\mathchar12349\nobreak\discretionary{}%
 {\usefont{OT1}{cmr}{m}{n}\char61}{}}
\endgroup
\def\cdot{\mathchar8705\nobreak\discretionary{}%
 {\usefont{OMS}{cmsy}{m}{n}\char1}{}}
\def\times{\mathchar8706\nobreak\discretionary{}%
 {\usefont{OMS}{cmsy}{m}{n}\char2}{}}
\AtBeginDocument{\mathcode`\==32768 \mathcode`\+=32768
\mathcode`\-=32768 }

\def\newbiblioname {\large Библиографические ссылки}

\def\eqalign#1{\null\,\vcenter{\openup\jot\m@th
  \ialign{\strut\hfil$\displaystyle{##}$&$\displaystyle{{}##}$\hfil
      \crcr#1\crcr}}\,}
\def\eqalignleft#1{\null\,\vcenter{\openup\jot\m@th
  \ialign{\strut$\displaystyle{##}$\hfil&$\displaystyle{{}##}$\hfil
      \crcr#1\crcr}}\,}

\def\b#1:{{\bf#1}, }

\def\beq#1{\begin{equation}\label{#1}}
\def\eeq{\end{equation}}

\def\b{\begin{eqnarray}}
\def\earr{\end{eqnarray}}

\def\la{\mathrel{\mathpalette\fun <}}
\def\ga{\mathrel{\mathpalette\fun >}}
\def\fun#1#2{\lower3.6pt\vbox{\baselineskip0pt\lineskip.9pt
\ialign{$\mathsurround=0pt#1\hfil##\hfil$\crcr#2\crcr\sim\crcr}}}

\def\b{\begin{eqnarray}}
\def\earr{\end{eqnarray}}

\def\la{\mathrel{\mathpalette\fun <}}
\def\ga{\mathrel{\mathpalette\fun >}}
\def\fun#1#2{\lower3.6pt\vbox{\baselineskip0pt\lineskip.9pt
\ialign{$\mathsurround=0pt#1\hfil##\hfil$\crcr#2\crcr\sim\crcr}}}

\def\la{\mathrel{\mathchoice {\vcenter{\offinterlineskip\halign{\hfil
$\displaystyle##$\hfil\cr<\cr\sim\cr}}}
{\vcenter{\offinterlineskip\halign{\hfil$\textstyle##$\hfil\cr
<\cr\sim\cr}}}
{\vcenter{\offinterlineskip\halign{\hfil$\scriptstyle##$\hfil\cr
<\cr\sim\cr}}}
{\vcenter{\offinterlineskip\halign{\hfil$\scriptscriptstyle##$\hfil\cr
<\cr\sim\cr}}}}}
\def\ga{\mathrel{\mathchoice {\vcenter{\offinterlineskip\halign{\hfil
$\displaystyle##$\hfil\cr>\cr\sim\cr}}}
{\vcenter{\offinterlineskip\halign{\hfil$\textstyle##$\hfil\cr
>\cr\sim\cr}}}
{\vcenter{\offinterlineskip\halign{\hfil$\scriptstyle##$\hfil\cr
>\cr\sim\cr}}}
{\vcenter{\offinterlineskip\halign{\hfil$\scriptscriptstyle##$\hfil\cr
>\cr\sim\cr}}}}}

\makeatletter
\renewcommand{\footnoterule}{\kern-3\p@
 \hrule width .4\columnwidth
 \kern 2.6\p@}
\renewcommand{\@makefnmark}{}
\renewcommand{\@makefntext}[1]{\hspace{2em}\hbox{\hss#1}}

\addto\captionsrussian{%
\def\abstractname{}
}

\renewcommand{\@biblabel}[1]{#1.}
\makeatother

\newcommand{\pud}{\hbox to 0.7em {\hspace{0.2em}.\hss $^d$}}
\newcommand{\pus}{\hbox to 0.4em {\hspace{0.01em}$''$\hss .}}
\newcommand{\pum}{\hbox to 0.7em {\hspace{0.2em}.\hss $^m$}}
\newcommand{\pug}{\hbox to 0.4em {\hspace{0.05em}.\hss $^{\circ}$}}
\newcommand{\Rim}[1]{\uppercase\expandafter{\romannumeral#1}}

\newcommand{\bnl}[1]{\begin{equation}\label{#1}}
\newcommand{\ed}{\end{equation}}

\makeatletter
\newcommand{\l@mytoc}[2]{\hbox to\textwidth%
        {\hspace{.5cm}\parbox[b]{15cm}{#1}}}
\makeatother

\begin{document}

\selectlanguage{russian}

\input boley.tex

\end{document}

%% file: boley.tex
\bibliographystyle{Kourovkastyle_}
\par
\hspace{0.0\textwidth}
\parbox[t]{0.95\textwidth}
{
\raggedleft
{\bf П.~Э.~Боли}\\
Институт радиоастрономии общества Макса Планка (MPIfR), Германия\\
\raggedright }

\footnote{\copyright\ Боли П.~Э., 2014}
\par
\medskip
\begin{center}
\bf
ИНФРАКРАСНАЯ ИНТЕРФЕРОМЕТРИЯ\\ И ИЗУЧЕНИЕ МАССИВНЫХ МОЛОДЫХ\\ ЗВЁЗДНЫХ
ОБЪЕКТОВ
\end{center}



\addcontentsline{toc}{subsection}{{\bf Боли П.~Э.} Инфракрасная
интерферометрия и изучение массивных молодых звёздных объектов}
\par
\medskip

\begin{minipage}{0.9\textwidth}
{\small Излагаются введение в теорию и практику  инфракрасной
интерферометрии на примере изучения массивных молодых звёздных объектов, основы интерферометрии, а также наблюдаемые параметры и критерии их выбора.  Обсуждаются последние достижения в
интерферометрических исследованиях массивных молодых звёздных
объектов.}\end{minipage} \par \vspace{2mm}

\selectlanguage{english}

\begin{minipage}{0.9\textwidth}
{\small An introduction to the theory and practical aspects of
infrared interferometry is given in the context of the study of
massive young stellar objects.  Basic interferometric concepts, as
well as observable quantities and their use, are presented.  Recent
advancements in interferometric studies of massive young stellar
objects are discussed.}\end{minipage}

\selectlanguage{russian}

\section*{Введение}


Звёзды с массой $\ga8$~M$_\odot$ (такие звёзды в данной обзорной лекции будут называться
<<массивными>>) составляют менее 0.4~\%
от общего числа всех звёзд, сформированных в настоящую эпоху
Вселенной.  Несмотря на это, по разным оценкам, считается, что при
рождении нового <<населения>> звёзд примерно 17~\% всеобщей звёздной
массы приходится именно на них~\cite{Kroupa01}.  Массивные звёзды
очень сильно влияют как на ближайшую окрестность, так и на содержащие
их галактики.

Подавляющее число массивных звёзд формируются в звёздных
скоплениях~\cite{deWit04,deWit05}, образовавшихся в плотных
молекулярных облаках.  Из-за своих высоких температур и светимостей
массивные звёзды способны ионизовать огромные объемы в окружающей
межзвёздной среде.  Это сильное излучение может даже ионизовать
плотные диски вокруг соседних звёзд, результатом чего может служить
частичное или полное разрушение околозвёздных
дисков~\cite{ODell93,Fang09}.  В то же время расширение ионизованной
зоны разгоняет окружающий газ, что может довести до рассеивания
родительского молекулярного облака и, следовательно, приостановить
дальнейшее образование звёзд в районе.

Б\'ольшую часть времени своего существования ($\sim10-30$~млн~лет)
массивные звёзды превращают водород в гелий в своих недрах за счёт
процесса нуклеосинтеза.  После ухода звезды с главной последовательности
образуются более тяжёлые элементы вплоть до железа.
Образование \emph{еще} более тяжёлых элементов происходит, когда
исчерпываются запасы ядерного топлива и массивная звезда взрывается
как сверхновая.  Этот великолепный взрыв разбрасывает материал в
межзвёздную и межгалактическую среды, таким образом обогащая их
тяжёлыми элементами, которые очень важны для жизни и ряда физических
процессов (в том числе и для радиативного охлаждения при коллапсе
молекулярных облаков и протозвёздных ядер).


Несмотря на важность роли, которую играют массивные звёзды в звёздных
скоплениях и галактической эволюции в целом, о механизмах их
образования и формирования известно довольно мало.  Основным
препятствием в образовании массивной звезды раньше считалось то,
что (при сферической симметрии) давление излучения на падающее на
звезду вещество должно превзойти силу тяготения довольно рано в
процессе коллапса протозвёздного ядра, таким образом приостановив
коллапс и ограничив массу звезды~\cite{Kahn74,Wolfire87}.  Однако эта
проблема решается в теоретических расчётах, если коллапс имеет не
сферический, а дискообразный вид~\cite{Yorke02,Krumholz07,Kuiper12}.
Тем не менее эти диски наблюдаются вокруг массивных звёзд с
трудом (если они вовсе наблюдаются, см.~ниже), и сказать
что-нибудь точно о них с точки зрения наблюдений пока крайне сложно (более подробное обсуждение теоретической стороны этой проблемы см.~в работе R.~Kuiper~\cite{Kuiper09}).

В лекции речь пойдёт о так называемых массивных молодых
звёздных объектах (MYSOs~--- Massive Young Stellar Objects~--- в
англоязычной литературе).  Данная классификация установлена
наблюдательным путём и не имеет единого определения, но в ходе
лекции ими будут условно считаться <<глубоко погружённые объекты со
светимостью $\ga10^4$~L$_\odot$>>.  \pagebreak Примечательно, что масса в это
определение напрямую не входит и что политкорректный термин
<<звёздный объект>> просто скрывает наше великое невежество, когда речь идёт о таких
объектах$^{1}.$\footnote{\hspace{-1.5mm}$^{1}$Например, каков спектральный класс объекта? Происходят ли в нем }
\footnote{ядерные реакции? Не на главной последовательности
ли он, и как это} \footnote{определить? И так далее.}

Наблюдения MYSOs осложняются двумя неизбежными фактами.  Во-первых,
из-за чрезмерно быстрой эволюции массивные звёзды уже находятся на
главной последовательности к тому моменту, когда они становится видны
в оптическом диапазоне (до этого они погружены в плотный газ
молекулярного облака и/или околозвёздной оболочки, где поглощение
света большое~--- $A_V \approx 10$---$100$~звёздных величин).  Поэтому
наблюдения MYSOs в оптическом диапазоне невозможны.  Во-вторых,
массивные звёзды и так по существу своему редки, что, как уже сказано, ведёт к тому, что они в среднем расположены на относительно
больших расстояниях от нас ($\ga1$~кпк).  Следовательно, в стандартных
режимах наблюдений на единичных телескопах с диаметром вплоть до
$\sim10$~м MYSOs не разрешаются из-за дифракционного предела: они
выглядят как точечные объекты.

Тем не менее в последнее десятилетие был осуществлён ряд работ, в
которых тем или иным наблюдательным способом удаётся обойти
дифракционный предел.  Особое место занимает инфракрасная
интерферометрия, в которой одновременно используются несколько
телескопов.  \emph{Эффективный} диаметр таких интерферометров
сегодня составляет десятки или сотни метров, что позволяет
достигать пространственного разрешения до $\sim0.001''$.
В то же время эти интерферометры работают на длинах волн
$\sim1$---$13$~мкм (в зависимости от прибора), где поглощение света за
счёт оболочки гораздо слабее, чем в оптическом диапазоне.

\section*{Астрономическая интерферометрия}

\subsection*{Интерферометрическая видность}

Одним из важнейших характеристик любого астрономического измерения
является его \emph{угловое разрешение}, которое в общем случае
зависит от длины волны, применяемых измерительных приборов,
атмосферных условий и т.~д.
\pagebreak

В области астрономии интерферометрия~--- это режим наблюдения, при
котором когерентно складываются вместе сигналы нескольких телескопов
(или нескольких апертур одного телескопа).  Достигнутое таким образом
угловое разрешение зависит от расстояния между телескопами
(так называемого базиса) и может на порядки превысить разрешение отдельных
составляющих интерферометр телескопов.  Первые интерферометрические
астрономические наблюдения были проведены для измерения диаметров
ближайших звёзд Майкельсоном и Писом в 1921~г.~\cite{Michelson21}.

Самой главной величиной любых интерферометрических наблюдений является
интерферометрическая видность $V$.  Эта величина включает в себя фазу
и интенсивность интерференционной картины, и, согласно теореме Ван
Циттерта---Цернике, равна преобразованию Фурье распределения
интенсивности $I$:
\begin{equation}
  \label{eq_Vdef}
V(u,v) = \frac{\int\int I(\alpha, \beta) e^{-2\pi i (u \alpha + v
    \beta)}\, d\alpha\, d\beta}{\int \int I(\alpha,
  \beta)\, d\alpha\, d\beta},
\end{equation}
где $\alpha$, $\beta$~--- угловые координаты (например, в угловых
секундах); $u$, $v$~--- угловые частоты (в обратных единицах).  Таким
образом, если при помощи интерферометра измерить $V$ (или часть этой
величины, например её амплитуду или фазу) для конечной выборки
пространственных частот, мы получим некую информацию о распределении
интенсивности $I$ (или, простыми словами, как выглядит наш
астрономический объект).  Распределение измерений $V$ по
$uv$-пространству называется $uv$-покрытием. Чем лучше
$uv$-покрытие, тем вернее наши представления о распределении
интенсивности $I$.  Следует отметить, что в пределе бесконечно многих
измерений видности $V$ на всех пространственных частотах $(u,v)$
функцию $I(\alpha,\beta)$ можно получить в полном виде посредством
обратного преобразования Фурье.

\subsection*{Пример использования интерферометрической видности}

\begin{figure}[t]
  \begin{center} \includegraphics[width=76mm]{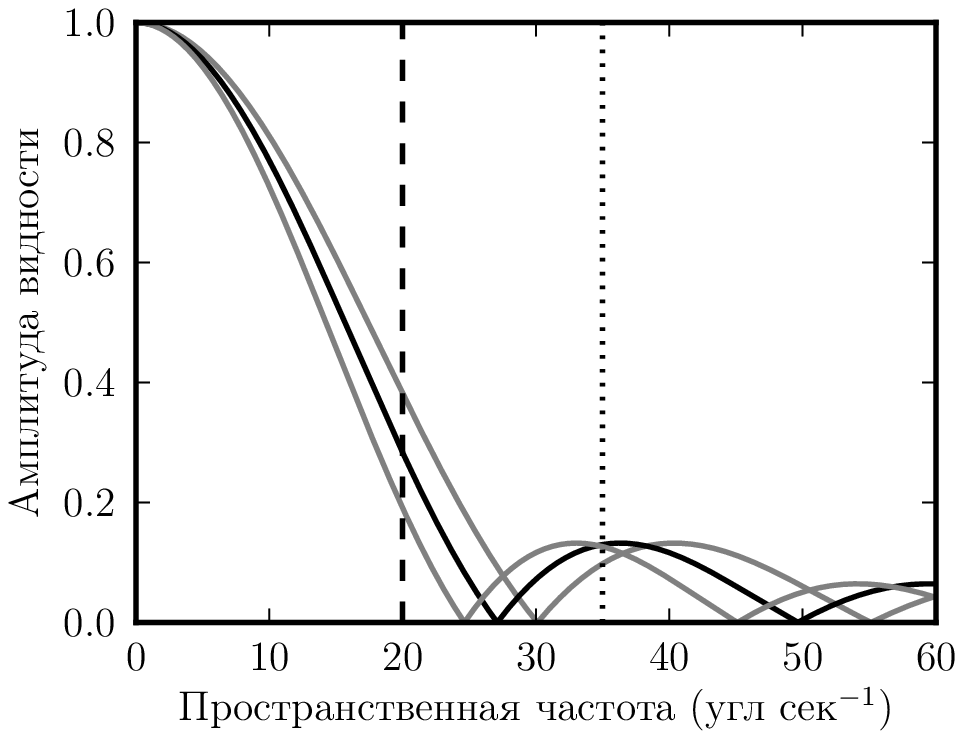}
  \caption{Функция видности однородного диска с диаметром $0.047''$ (чёрная кривая) и аналогичные функции для диска с
          диаметром на 10~\% меньше и больше (серые кривые).
          Штрихованной линией указана одна частота, на которой
          достаточно легко различить эти три кривые и
          соответственно определить диаметр звезды в данном примере.
          Точечной линией указана другая частота, на которой гораздо
          сложнее однозначно определить диаметр
          звезды}  \label{fig_diameter} \end{center}
\end{figure}

Вышеизложенное соотношение интерферометрической видности с
распределением интенсивности довольно абстрактное для первого
ознакомления с интерферометрией.  В этом разделе приводится пример
использования наблюдений видности $V$ для измерения диаметра звезды.

В первом приближении наблюдаемую плоскость звезды можно описать как
однородный диск.  Функция видности такого диска имеет следующий
вид~\cite{Berger07}:
\begin{equation}
V(u, v) = 2 \frac{J_1(\pi \Theta \sqrt{u^2+v^2})}{\pi \Theta \sqrt{u^2+v^2}},
\end{equation}
где $\Theta$~--- угловой диаметр диска; $J_1$~--- функция Бесселя
первого рода и первого порядка.  Так как функция видности зависит от
всего одного параметра (диаметра диска), измерение видности на одной
лишь частоте \emph{может быть} (в зависимости от частоты) достаточно,
чтобы определить звёздный диаметр.  На рис.~\ref{fig_diameter}
показана функция видимости диска с диаметром $0.047''$ (это
диаметр Бетельгейзе, измеренный Майкельсоном и Писом в
1921~г.~\cite{Michelson21}) вместе с функциями видности для диска с
диаметром на 10~\% больше и меньше того значения.  Очевидно, что эти
кривые пересекаются в некоторых местах, и возможность их различить
зависит от того, на какой пространственной частоте проводятся
наблюдения (что зависит в первую очередь от базиса~--- расстояния
между телескопами).  Хотя, если хорошо подобрать пространственную
частоту или проводить наблюдения на нескольких частотах, диаметр
звезды можно определить с большой точностью (следует отметить, что
\emph{правильность} этого определения всё-таки зависит от того,
насколько справедливо предположение о том, что звезда выглядит как
однородный диск).

Таким образом, если можно построить простую модель распределения
интенсивности интересующего нас объекта, то достаточно <<легко>> найти
параметры той модели.  С другой стороны, если объект имеет более
сложный вид или вообще невозможно заранее его предсказать, задача
намного осложняется; её решение обычно требует хорошего $uv$-покрытия.

\subsection*{Музыкальная аналогия}

\begin{figure}[t]
\centering
\includegraphics[width=90mm]{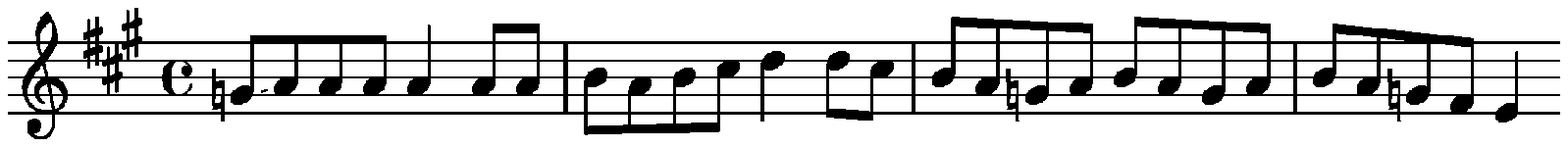}
\includegraphics[width=90mm]{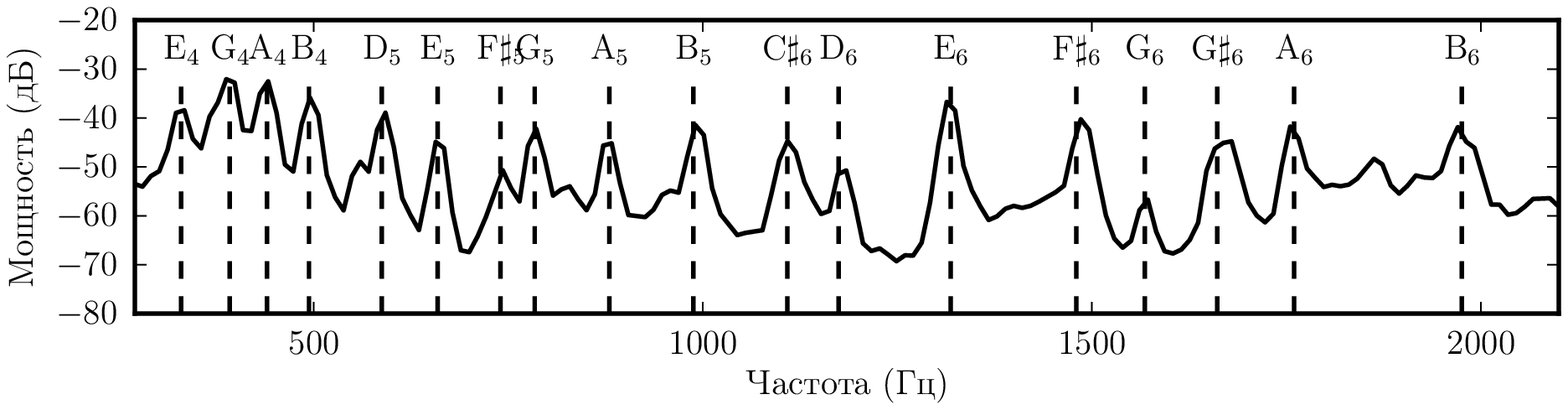}
\caption{Нотная запись открывающей фразы песни <<Salt
Creek>> (сверху). Спектральное распределение мощности звуковой
записи данной фразы, воспроизведенной на мандолине.  Штриховыми
линиями указаны частоты отдельных музыкальных звуков (снизу)}
\label{fig_saltcreek}
\end{figure}

Чтобы почувствовать суть задачи, когда исходная форма распределения
интенсивности $I$ полностью нам неизвестна, рассмотрим музыкальную
аналогию.  На рис.~\ref{fig_saltcreek} представлены нотная запись
короткой фразы из песни <<Salt Creek>> (сверху) и спектральное
распределение мощности (СПМ) звуковой записи той же фразы.  СПМ равно
амплитуде преобразования Фурье среднего по времени звукового сигнала,
поэтому спектр мощности можно считать аналогом функции видности.

Сейчас представим, что мы умеем измерять СПМ только на отдельных
(звуковых) частотах.  На рис.~\ref{fig_saltcreek} видно, что повышения
мощности приходятся на соответствующие музыкальным тонам частоты,
поэтому по дискретным измерениям спектра мощности достаточно легко
судить об относительной важности различных нот.  Если хорошо подобрать
$10$---$20$ частот (что напоминает реальную ситуацию в инфракрасной
интерферометрии), то заметим, что повышенная мощность наблюдается на
частотах, соответствующих нотам A, B, C$\sharp$, D, E, F$\sharp$ и
G$\sharp$.  Отсюда следует, что исходный <<сигнал>> воспроизводился в
тональности или ля мажор, или фа-диез минор.  Если постараться, то можно заметить маленький пик на частоте ноты G$_6$, что говорит о том,
что лад данного воспроизведения не мажорный и не минорный, а на самом
деле миксолидийский.

Такой анализ спектра мощности даёт очень тонкую информацию об исходной
звуковой записи~--- даже при сравнительно небольшом количестве
измерений.  То, что исходный звуковой сигнал, в общем, музыкальный, позволяет сразу получить его
тональность и лад.  Имея опыт в анализе
преобразования Фурье музыкальных произведений, нетрудно представить,
что мы сможем в будущем различать, например, какие музыкальные
инструменты использовались, в каком они состоянии, какого жанра композиция и какие этнические элементы в ней присутствуют\ldots{}  Но такой анализ, увы, нам никогда не скажет, как песня \emph{звучит}.  Так же как и в
астрономической интерферометрии, наблюдения функции видности не даёт нам \emph{картину} (хотя существует несколько способов для <<реконструкции>> картин из интерферометрических наблюдений).

\section*{Применение инфракрасной интерферометрии\\ для изучения
массивных молодых\\ звёздных объектов}

Сегодня (или в недавнем прошлом) в мире действуют несколько
инфракрасных интерферометров, таких как ISI~\cite{Hale00},
CHARA~\cite{tenBrummelaar05}, Keck-I~\cite{Tuthill00} (демонтирован).
Но подавляющее большинство интерферометрических работ по массивным
молодым звёздным объектам на инфракрасных длинах волн было сделано на
VLTI Европейской южной обсерватории на приборах AMBER ($1$---$2.4$~мкм, три телескопа)~\cite{Petrov07} и MIDI ($8$---$13$~мкм, два телескопа)~\cite{Leinert03}.

\subsection*{Первые шаги}

Работы по массивным молодым звёздным объектам с использованием
инфракрасных интерферометров начали публиковаться только в 2007~г.
В одной из первых работ de~Wit и др.~\cite{deWit07} представили
амплитуду видности в полосе $N$ ($8$---$13$~мкм) для одной точки в
$uv$-пространстве для объекта W33A (расстояние 3.8~кпс,
светимость $\sim10^5$~L$_\odot$).  В отличие от представленного
выше примера со звёздным диаметром, в котором предполагалось, что
исходная форма распределения интенсивности имеет вид однородного
диска, тут форма распределения интенсивности заранее неизвестна.
Поэтому одинарная точка в $uv$-пространстве, кажется, не так уж показательна.  Однако благодаря именно этим интерферометрическим наблюдениям впервые для молодой звезды были установлены наличие
<<тёплого>> ($\sim300$~К) вещества и его примерные размеры
($100$---$200$~а.~е.).

\begin{figure}[ht!]
\centering
\includegraphics[width=0.90\textwidth]{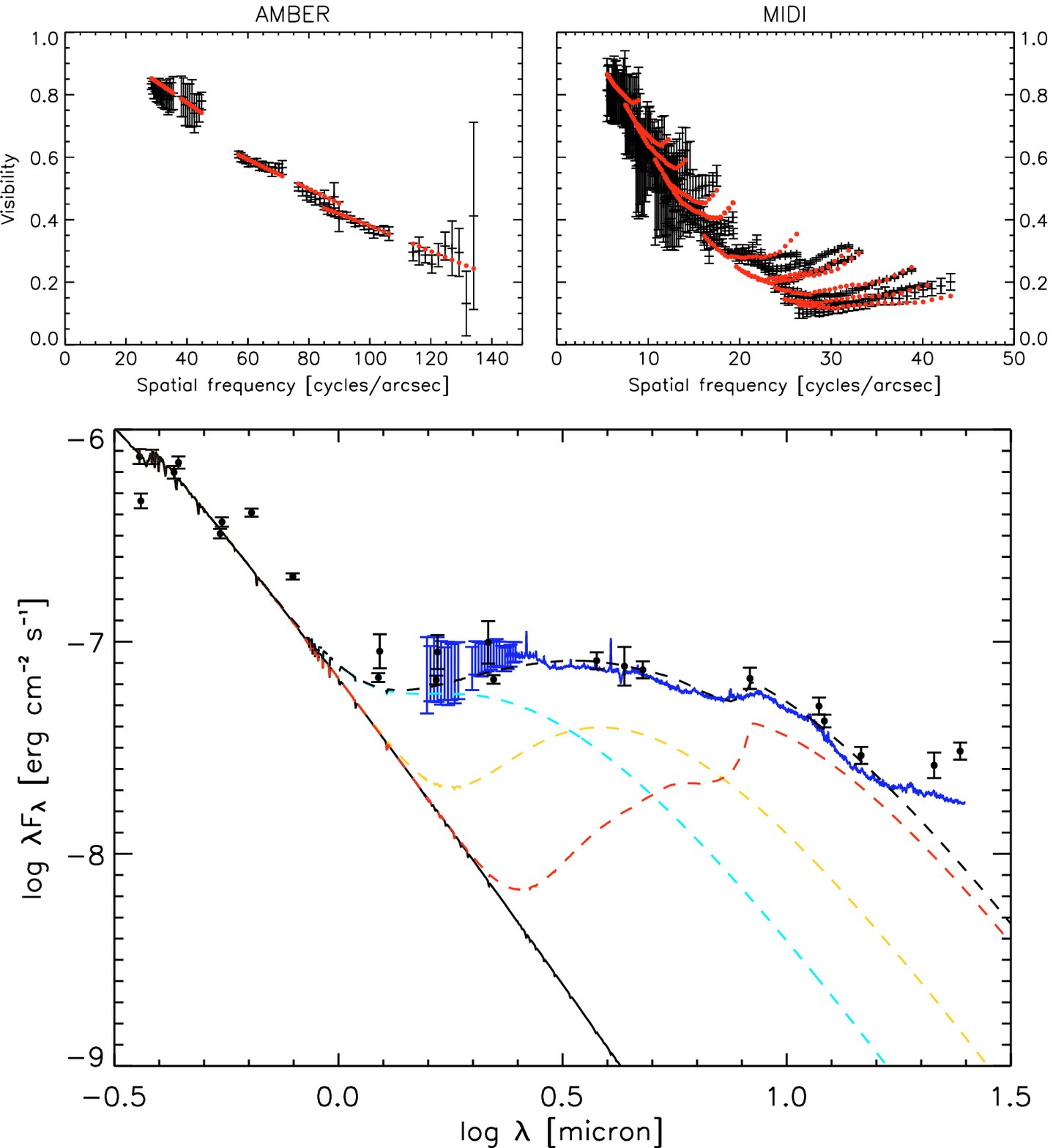}
\caption{Видность в ближнем (слева) и среднем (справа)
инфракрасном диапазоне для объекта MWC~297.  Наблюдения показаны с
усами; модельные значения показаны точками (сверху).  Спектральное распределение энергии.  Верхняя точечная линия изображает суммарное излучение модели (снизу).  Рисунок из
работы~\cite{Acke08}}
\label{fig_Acke}
\end{figure}

Более сложное моделирование интерферометрических наблюдений объекта
MWC~297 (расстояние 250~пк, светимость $\sim10^4$~L$_\odot$)
в~полосах $H$ ($1.6$---$2.0$~мкм), $K$ ($2.0$---$2.4$~мкм) и $N$ было сделано
авторами Acke и др.~\cite{Acke08}.  В этой работе используется простая
модель, составленная из гауссовых компонентов, которые излучают как
чёрные тела.  Данная модель хорошо описывает интерферометрические
данные и спектральное распределение энергии (рис.~\ref{fig_Acke}) и
указывает на наклонённую структуру с размером $\la1.5$~а.~е., при этом с
минимальным комплектом свободных параметров (их всего семь).

\subsection*{Интерферометрия и модели переноса излучения}

Начиная с 2009~г. начали публиковаться работы с общей методикой
использования сетки из 200\,000 моделей переноса излучения от Robitaille и
др.~\cite{Robitaille06}.  Такая тактика привлекательна тем, что она
позволяет посчитать видность и спектральное распределение энергии для
сложных распределений вещества, например в виде околозвёздного диска
и/или оболочки.  В
работах \cite{Linz09,deWit10,Follert10,Kraus10,deWit11,Grellmann11}
эти относительно сложные модели использовались для подгонки
спектрального распределения энергии; интерферометрические наблюдения в
основном применялись только для того, чтобы исключить неподходящие
модели.  К сожалению, из-за большого количества свободных параметров и
высокого вырождения в данном подходе остаётся место для сомнения в
интерпретации интерферометрических наблюдений массивных молодых
звёздных объектов.
\pagebreak
\subsection*{Первое интерферометрическое изображение\\ массивного
молодого звёздного объекта}

\begin{figure}[ht]
\centering
\includegraphics[width=0.6\textwidth]{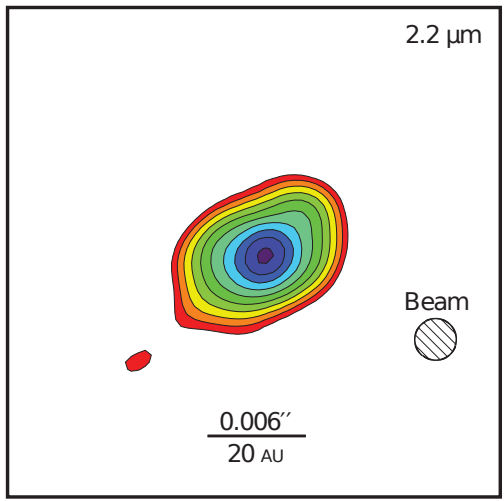}
\caption{Первое реконструированное интерферометрическое изображение
массивного молодого звёздного объекта IRAS~13481-6124.  Рисунок
 из работы~\cite{Kraus10}}
\label{fig_Kraus}
\end{figure}

Очень важное развитие в изучении массивных молодых звёзд было сделано
авторами Kraus и др.~\cite{Kraus10} в 2010~г.  Благодаря
превосходному заполнению $uv$-плоскости им удалось получить первое
реконструированное изображение объекта IRAS~13481-6124 (расстояние
3.6~кпк, светимость $\sim3$---$6\times10^4$~L$_\odot$) на длине волны
2.2~мкм.  Методика реконструкции изображений очень хороша тем, что она
даёт независимое от моделей изображение (в отличие от использованных в
предыдущих двух разделах приёмов). Однако её можно использовать только
тогда, когда имеется отличное заполнение $uv$-плоскости, что требует много
наблюдательного времени.

Полученное изображение (рис.~\ref{fig_Kraus}) показывает вытянутую
структуру, которая перпендикулярна крупномасштабному молекулярному
истечению.  Это пока единственное изображение массивного молодого
звёздного объекта, в котором разрешается излучение от тёплой
околозвёздной пыли на масштабе десятков астрономических единиц.
Авторы  работы считают, что показанная на рис.~\ref{fig_Kraus}
структура суть околозвёздный диск.  Однако эта гипотеза была
сделана на основе подобранных моделей переноса излучения (см. выше),
стоит рассмотреть другие возможности (Boley, Kraus и др., готовится
к печати). \pagebreak

\subsection*{Интерферометрический обзор\\ массивных молодых звёздных
объектов}

Все вышеперечисленные работы были сосредоточены на изучении отдельных
объектов.  Общее количество рассмотренных таким образом объектов не
превышает десяти, при этом в каждой работе формируются различные допущения
и применяются различные подходы.  В связи с этим авторами Boley и
др.~\cite{Boley13} был представлен интерферометрический обзор в полосе
$N$ для выборки из 25 объектов высокой и средней массы.  Относительно
простой геометрический анализ в этой работе избегает излишних
сложностей, чтобы претендовать на более или менее однородное
рассмотрение совокупности наблюдений.

\begin{figure}[ht!]
\centering
\includegraphics[width=0.95\textwidth]{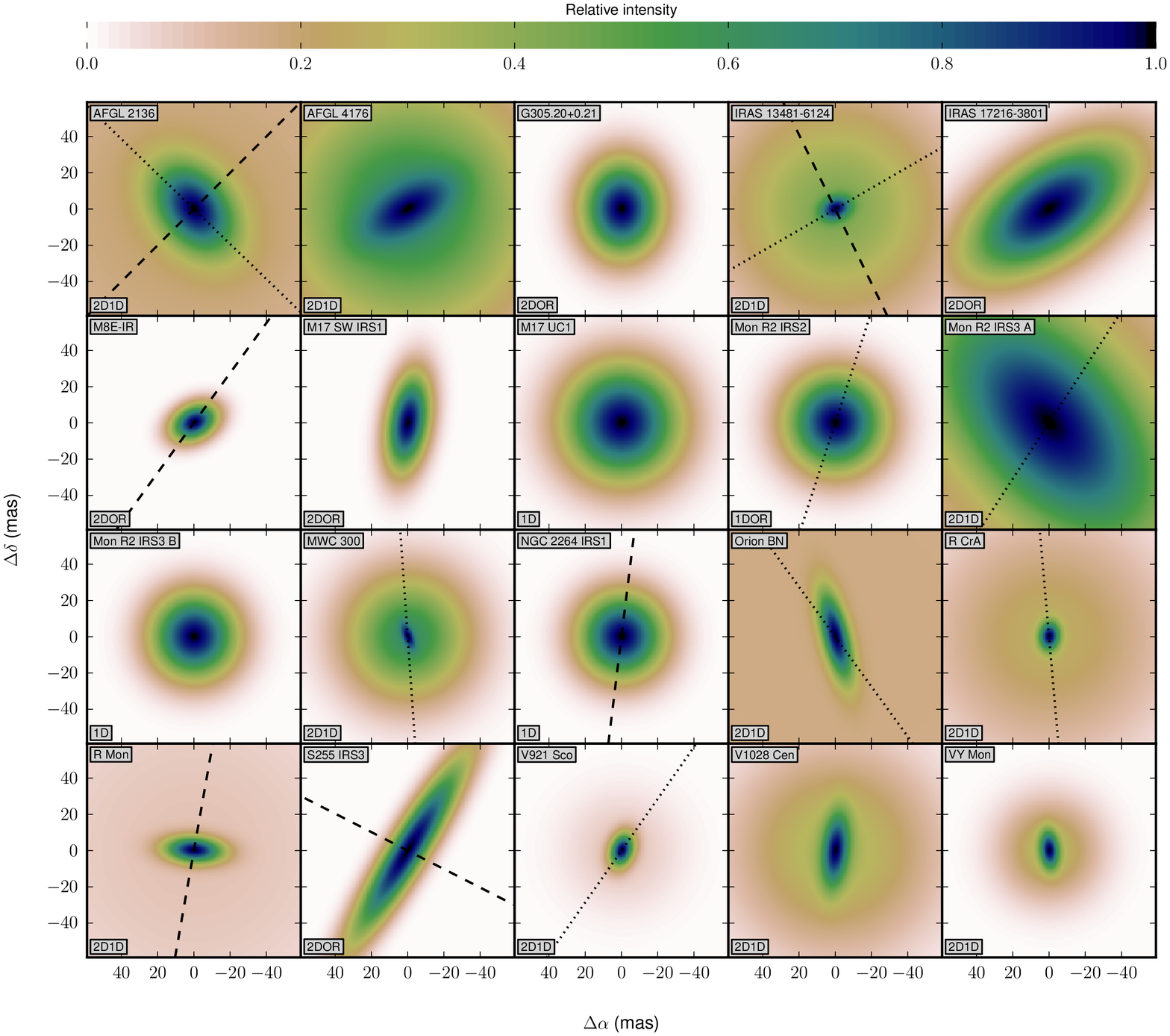}
\caption{Результаты подгонки гауссовых моделей к интерферометрическим
наблюдениям для двадцати молодых звёздных объектов высокой и средней массы.
Точечными линиями показаны позиционный угол диска (если он есть),
штрихованными~--- позиционный угол истечения (если он есть).
Подписями снизу указано, какая модель использовалась~--- в зависимости
от $uv$-покрытия~--- для подгонки.  Рисунок взят из
работы~\cite{Boley13}}
\label{fig_Boley}
\end{figure}

На рис.~\ref{fig_Boley} показаны результаты подгонки одномерных
(<<1D>>, <<1DOH>>) и двумерных (<<2DOH>>, <<2D1D>>) гауссовых моделей
к интерферометрическим наблюдениям вместе с направлением истечений
или дисков (если есть о них информация в опубликованной литературе).
Размерность использованной гауссовой модели зависит от
$uv$-покрытия, когда возможно, используется двумерная модель.

Примечательно, что компактное излучение в среднем инфракрасном
диапазоне разрешалось для 20 из 25 объектов, что говорит о широком
распространении <<тёплого>> околозвёздного вещества на расстоянии
десятков астрономических единиц вокруг молодых массивных звёзд.
Дальше, когда $uv$-покрытие допускает использование двумерной модели,
как правило, наблюдается значительное отклонение от сферической
симметрии.  Чаще всего это компактное излучение вытянуто
перпендикулярно направлению истечения или параллельно диску (например,
AFGL~2136, IRAS~13481-6124, R~Mon), хотя стоит подчеркнуть, что
иногда получается совсем наоборот (например, Mon~R2~IRS3A, M8E-IR).

\section*{Заключение}

Лекция была посвящена основам астрономической
ин\-тер\-фе\-ро\-мет\-рии в инфракрасном диапазоне.  Особое внимание
было уделено изучению массивных молодых звёздных объектов, однако \pagebreak
представленные методы также применимы и для изучения других объектов
(в том числе кратных систем, звёзд типа Хербига и Т~Тельца,
проэволюционировавших звёзд, галактик с активными ядрами, планетных
систем и т.~д.).

Благодаря инфракрасной интерферометрии впервые становится возможным
разрешение околозвёздного материала вокруг массивных звёзд на
актуальных для аккреционных процессов масштабах.  Это означает, что
именно сейчас открывается новая территория в области образования
массивных звёзд.  \pagebreak Первые наблюдения, хотя и сильно ограничены,
уже доставляют нам уникальную информацию об этих объектах.
Разумеется, в скором будущем можно ожидать развития как самих
интерферометрических приборов, так и методов анализа
интерферометрических данных.

Наконец, для заинтересованного читателя рекомендуются обзорные
статьи~\cite{Beuther07} (по формированию массивных звезд)
и~\cite{deWit12} (по использованию инфракрасной интерферометрии для
изучения молодых массивных звёзд).



\input journal.tex
\bibliography{boley}

%% file: journal.tex
\def\apj{Astrophys.~J}
\def\aatr{Astron.~Astroph.~Trans}
\def\aaps{Astron.~and Astrophys.~Suppl.~Ser}
\def\pasp{Publ.~Astron.~Soc.~Pac}
\def\gca{Geochim.~Cosmochim.~Acta}
\def\aap{Astron.~Astrophys}
\def\aspcs{ASP~Conf.~Ser}
\def\asrep{Astron.~Rep}
\def\nat{Nature}
\def\apjl{Astrophys.~J.~Lett}
\def\apjs{Astrophys.~J.~Suppl.~Ser}
\def\aj{Astron.~J}
\def\mnras{Mon.~Not.~R.~Astron.~Soc}
\def\araa{Ann.~Rev.~Astron.~Astrophys}
\def\jcp{J.~Chem.~Phys}
\def\apss{Astrophys.~Space.~Sci}
\def\prl{Phys.~Rev.~Lett}
\def\phrva{Phys.~Rev.~A}
\def\phlb{Phys.~Let.~B}
\def\pf{Phys.~Fluids}
\def\azh{Астрон.~журн}
\def\pazh{Письма~в~Астрон.~журн}
\def\jgr{J.~Geophys.~Res}
\def\cemda{Celest.~Mech.~Dyn.~Astr}
\def\jcoph{J.~Comp.~Phys}
\def\cophc{Comput.~Phys.~Commun}
\def\phpl{Physics~of~Plasmas}
\def\pasj{Publ.~Astron.~Soc.~Jpn}
\def\avest{Астрон.~вестн}
\def\jrasc{J.~R.~Astron.~Soc.~Can}
\def\cemec{Celest.~Mech}
\def\pasau{Proc.~Astron.~Soc.~Aust}
\def\puasau{Publ.~Astron.~Soc.~Aust}
\def\jasa{J.~Acoust.~Soc.~Am}
\def\jfm{J.~Fluid~Mech}
\def\cajph{Can.~J.~Phys}
\def\mitag{Mitt.~Astron.~Ges}
\def\bain{Bull.~Astron.~Inst.~Neth}
\def\epsl{Earth~Planet.~Sci.~Lett}
\def\ibvs{Inf.~Bull.~Variable~Stars}
\def\arep{Astr.~Rep}
\def\phr{Phys.~Rep}
\def\astl{Astron.~Letters}
\def\sci{Science}
\def\jqsrt{J.~Quant.~Spectrosc.~Radiat.~Transfer}
\def\emp{Earth,~Moon~and~Planets}
\def\icar{Icarus}
\def\pss{Planet.~Space~Sci}
\def\qjras{Q.~J.~R.~Astron.~Soc}
\def\nimpa{Nucl.~Instrum.~Methods~Phys.~Res.,~Sect.~A}
\def\soph{Sol.~Phys}
\def\lnm{Lect.~Notes~in~Math}
\def\an{Astron.~Nach}
\def\aph{Astroparticle~Physics}
\def\adspr{Adv.~Space~Res}
\def\geoj{Geophys.~J}
\def\caosp{Contrib.~Astron.~Obs.~Skalnat{\'e}~Pleso}
\def\vestcpbu{Вестн.~С.-Петерб.~ун-та}
\def\izvvrad{Изв.~вузов.~Радиофизика}
\def\izvans{Изв.~AH~CCCP}
\def\vestvgu{Вестн.~ВолГУ}
\def\vestsibgau{Вестн.~СибГАУ}
\def\bamass{Bull.~Am.~Astron.~Soc}
\def\rmxaa{Rev.~Mex.~Astron.~Astrofis}
\def\aapr{Astron.~Astrophys.~Rev}
\def\acp{Atmosphere~Chem.~Phys}
\def\cosiss{Космич.~исслед}
\def\ssrv{Space~Sci.~Rev}
\def\jmph{J.~Math.~Phys}
\def\rvmps{Rev.~Mod.~Phys.~Suppl}
\def\rvmp{Rev.~Mod.~Phys}
\def\prd{Phys.~Rev.~D}
\def\nuphs{Nuc.~Phys.~B~Proc.~Suppl}
\def\nuphb{Nuc.~Phys.~B}
\def\skytel{Sky~Telesc}
\def\thmc{Тез.~международ.~конф}
\def\mmc{Материалы~международ.~конф}
\def\mvrc{Материалы~всерос.~конф}
\def\cntc{Сб.~науч.~тр.~конф}
\def\tmnpc{Тр.~Международ.~науч.-практ.~конф}
\def\ctc{Сб.~тр.~конф}
\def\mcnctone{Тр.~31-й Международ.~студ.~науч.~конф., Екатеринбург, 28 янв.~--- 1~февр. 2002~г}
\def\mcnctto{Тр.~32-й Международ.~студ.~науч.~конф., Екатеринбург, 3---7 февр. 2003~г}
\def\mcncttr{Тр.~33-й Международ.~студ.~науч.~конф., Екатеринбург, 2---6 февр. 2004~г}
\def\mcnctfo{Тр.~34-й Международ.~студ.~науч.~конф., Екатеринбург, 31 янв.~--- 4~февр. 2005~г}
\def\mcnctfi{Тр.~35-й Международ.~студ.~науч.~конф., Екатеринбург, 30 янв.~--- 3~февр. 2006~г}
\def\mcnctsi{Тр.~36-й Международ.~студ.~науч.~конф., Екатеринбург, 29 янв.~--- 2~февр. 2007~г}
\def\mcnctse{Тр.~37-й Международ.~студ.~науч.~конф., Екатеринбург, 28 янв.~--- 1~февр. 2008~г}
\def\mcnctei{Тр.~38-й Международ.~студ.~науч.~конф., Екатеринбург, 2---6 февр. 2009~г}
\def\mcnctni{Тр.~39-й Международ.~студ.~науч.~конф., Екатеринбург, 1---5 февр. 2010~г}
\def\mcncforty{Тр.~40-й Международ.~студ.~науч.~конф., Екатеринбург, 31 янв.~--- 4~февр. 2011~г}
\def\mcncfone{Тр.~41-й Международ.~студ.~науч.~конф., Екатеринбург, 30 янв.~--- 3~февр. 2012~г}
\def\tmc{Тр.~Международ.~конф}
\def\tc{Тр.~конф}
\def\thc{Тез.~конф}

\def\IAUsympc{Proc.~IAU~Symp}
\def\IAUsymp{Proc.~IAU~Symp}
\def\IAUcoll{Proc.~IAU~Colloquia}
\def\prconf{Proc.~conf}
\def\princonf{Proc.~int.~conf}
\def\mvrnc{Материалы~всерос.~науч.~конф}